\documentclass[12pt]{article}
\usepackage{amsmath,amssymb}
 \usepackage{graphicx}

\begin{document}

\begin{center}

{\bf\large Eigenmodes of elastic vibrations of quaking
 neutron star encoded in QPOs on light curves of SGR flares}

\vspace{0.9cm}

{ Sergey Bastrukov$^{1,2}$,   Hsiang-Kuang Chang$^1$, Irina
Molodtsova$^2$, Gwan-Ting  Chen$^1$}

\vspace{0.7cm}

{\it \noindent $^1$ Department of Physics and  Institute of
Astronomy,
  National Tsing Hua University, Hsinchu, 30013, Taiwan \\

\noindent $^2$ Laboratory of Informational Technologies, Joint
Institute for Nuclear Research, \\ 141980 Dubna Russia

}

\end{center}

\begin{abstract}
 The Newtonian solid-mechanical theory of non-compressional spheroidal and torsional nodeless elastic vibrations
 in the homogenous crust model of a quaking neutron star is developed and applied to
 the modal classification of the quasi-periodic oscillations (QPOs) of X-ray luminosity in the aftermath
 of giant flares in SGR 1806-20 and SGR 1900+14.
 A brief outline is given of Rayleigh's energy method
 which is particular efficient when computing the frequency of nodeless elastic spheroidal and torsional shear modes
 as a function of multipole degree of nodeless vibrations and two input parameters -- the
 natural frequency unit of shear vibrations carrying information about equation of state and fractional depth
 of peripheral seismogenic layer. In so doing we discover that the dipole overtones of both spheroidal
 and torsional nodeless vibrations possess the properties of Goldstone soft modes.
 It is shown that
 obtained spectral formulae reproduce the early suggested identification
 of the low-frequency QPOs from the range $30\leq \nu \leq 200$ Hz as torsional
 nodeless vibration modes
 $\nu(_0t_\ell)$ of multipole degree $\ell$ in the interval $2\leq \ell \leq
 12$. Based on this identification, which is used to fix the above mentioned input parameters of derived spectral
 formulae, we compute the frequency spectrum of nodeless spheroidal
 elastic vibrations $\nu(_0s_\ell)$. Particular attention is given to the low-frequency QPOs in the data for
 SGR 1806-20 whose physical origin has been called into question.
 Our calculations
 suggest that these unspecified QPOs are due to nodeless dipole torsional and dipole spheroidal elastic shear
 vibrations: $\nu(_0t_1)=18$ Hz and
 $\nu(_0s_1)=26$ Hz.
\end{abstract}


\section{Introduction}
 The discovery of QPOs of X-ray
 luminosity in the aftermath of giant flares SGR 1806-20
 and SGR 1900+14 (Israel et al 2005; Strohmayer \& Watts 2006; Israel 2007; Watts \& Strohmayer 2007)
 with concomitant interpretation of QPOs as caused by quake-induced differentially rotational seismic vibrations
 has stimulated remarkable developments in the magnetar asteroseismology (e.g. Piro 2005,
 Samuelsson \& Andersson 2007a,
 2007b; Lee 2007; Levin 2007; Watts \& Reddy 2007; Sotani et al 2007; Bastrukov et al 2007a and references therein).
 Following the above interpretation and presuming the dominant role of the elastic restoring force, the focus
 of  most theoretical works is on computing the frequency spectra of
 odd-parity torsional mode of shear vibrations and less attention is paid to the even parity spheroidal
 elastic mode. However, from the viewpoint of modern global seismology (Lay \&
 Wallace 1995; Aki \& Richards 2003), the spheroidal vibrational mode in a solid star and planet
 has the same physical significance as the toroidal one in the sense
 that these two fundamental modes owe their existence to one and the same restoring
 force (e.g. McDermott, Van Horn, Hansen 1988; Bastrukov, Weber, Podgainy
 1999, Bastrukov et al 2007b). In this light there is a possibility
 that, by not considering both these modes on an equal footing, we may
 miss discovering certain essential novelties which are consequences of solid mechanical laws governing
 seismic vibrations of superdense matter of neutron stars.
 Adhering to this attitude and continuing the investigations recently reported by Bastrukov et al (2007a),
 we derive here spectral equations for the frequency of both spheroidal and torsional elastic nodeless vibrations
 in the solid crust of quaking neutron star and examine what conclusions can be drawn regarding
 low-frequency QPOs whose physical nature still remain unclear.
 In Sec.2 by use of the energy variational method we derive spectral formulae for the frequencies of
 the nodeless spheroidal and torsional elastic
 vibrations locked in the finite-depth seismogenic layer. Particular
 attention is given to the dipole spheroidal and
 torsional vibrations possessing properties of Goldstone's soft modes. In sec.3, the obtained spectral
 formulae are applied to a modal analysis of available data on the above mentioned QPOs.
 The obtained results are highlighted in Sec.4.

 \section{Frequency of nodeless spheroidal and torsional elastic shear vibrations in homogeneous crust}

 In this paper we follow the line of argument of the standard two-component, core-crust, model of quaking neutron star
 (Link, Franco, Epstein 2000) in which metal-like material (composed of nuclei dispersed in the sea of relativistic
 electrons) of the finite-depth seismogenic crust is treated as
 a highly robust to compressional distortions elastic continuous medium of a uniform density
 $\rho$ and characterized by constant value of shear modulus $\mu$. Also, the model presumes that
 the quake-induced seismic vibrations driven by bulk force of pure shear elastic deformations,
 which are not accompanied by fluctuations in density $\delta \rho=-\rho\,\nabla_k u_k=0$,
 can be modeled equation
 of Newtonian, non-relativistic, solid mechanics
 \begin{eqnarray}
 \label{e2.1}
&& \rho{\ddot u}_i=\nabla_k\sigma_{ik},\quad \sigma_{ik}=2\,\mu\,
u_{ik},\quad u_{ik}=\frac{1}{2}[\nabla_i u_k+\nabla_k u_i],\quad
u_{kk}=\nabla_k u_k=0.
 \end{eqnarray}
 From now on $u_i({\bf r},t)$ stands for the field of material displacements in the crust
 of the depth $\Delta R=R-R_c$ with $R$ and $R_c$ being radii of star and core, respectively.
 The linear relation between tensors of shear elastic stresses $\sigma_{ik}$ and
 shear deformations or strains $u_{ik}$ is the
 Hooke's law of elastic (reversal) shear deformations.
  whose strength is characterized by shear modulus

 In what follows we focus on poorly investigated regime of quasistatic shear vibrations whose mathematical analysis
 is quite different form vibrations in the regime of standing waves (Bastrukov at al 2007b).
 In the latter case, the solution of eigenfrequency problem consists in searching for
 the wave numbers $k$ of standing waves of material displacements ${\bf u}$ obeying the vector
 Helmholtz equation $\nabla^2 {\bf u}+k^2{\bf u}=0$ supplemented by two boundary conditions on the edges of
 seismogenic zone, that is, on the core-crust interface and the star surface\footnote{It is appropriate
 to note that the very problem of vibrational
 modes and its solutions for
 the case of standing wave in entire volume of homogeneous elastically deformable solid sphere has first been
 tackled and solved by Lamb \cite{LAMB-H}. The application of Lamb's theory to analysis
 of seismic vibrations of the solid Earth model is extensively discussed in monographs (Jeffreys
 1976; Lapwood \& Usami 1981).}. The two fundamental
 (orthogonal and different in parity) solutions to the vector Helmholtz equation one of which
 is given by the positive parity poloidal (polar) field and another by negative parity toroidal
 (axial) field provide a basis for generally accepted Lamb's classification of vibrational modes
 as spheroidal or $s$-mode (in which the field of displacement is described by poloidal
 field) and torsional or $t$-mode (in which the field of displacement is described by poloidal
 field). The substitution of these fundamental solutions in
 boundary conditions, which are motivated by physical arguments, leads to the coupled system of transcendent
 dispersion equations for the wave vector $k$ having the form of bilinear combinations of spherical Bessel and Neumann
 functions with highly involved distribution of nodes along the radial coordinate of the layer of thickness $\Delta R$.
 It is the roots of these dispersion equations yield the discrete set of the wave numbers $k$ uniquely related
 to the frequency of oscillations: $\omega=k\,c_t$.
 The systematic application of the above method to the case of torsional elastic modes trapped in the
 homogeneous crust is presented in  (Bastrukov et al 2007a) and
 pointed out that similar procedure holds true for the case of spheroidal vibrational mode
 in the homogeneous crust model.

 The situation is quite different in the case of long wavelength vibrations, that is, when wave vector
 $k=(2\pi/\lambda)\to 0$ and, hence, the wavelength $\lambda\to \infty$. This limit corresponds to regime
 of quasistatic, substantially nodeless, vibrations in which the fields of oscillating
 material displacement subject to the vector Laplace equation (Bastrukov et al 2007b)
  \begin{eqnarray}
  \label{e2.2}
 && \nabla^2 {\bf u}({\bf r},t)=0.
 \end{eqnarray}
 Understandably, this last equation can be thought of as the long wavelength limit of vector Helmholtz
 equation describing standing-wave regime of vibrations (Bastrukov et
 al 2007a; 2007b). Our interest to the regime of nodeless seismic vibrations is motivated by
 arguments of works (Samuelsson \& Andersson 2007a; 2002b) in which based on equations of general relativity
 it was found that low-frequency QPOs in the above mentioned SGR's flare can be identified with
 low-$\ell$ nodeless torsional elastic vibrations locked in the crust. And it is somewhat surprising that
 no such problem has hitherto been properly analyzed on the basis
 of equations of Newtonian, non-relativistic, solid mechanics.
 To this end, in (Bastrukov et al 2007b) it has been shown for the first time that the problem of
 computing frequency spectra of both spheroidal and torsional
 modes of global nodeless vibrations of the solid star can be uniquely solved with aid of the
 energy method which is due to Rayleigh. In the present paper this method is extended to
 spheroidal mode of nodeless elastic vibrations which are considered in one line with torsional mode.

 The stating point of the energy variational method is the integral
 equation of the energy balance
\begin{eqnarray}
 \label{e2.3}
  \frac{\partial }{\partial t}\int \frac{\rho {\dot u}^2}{2}\,d{\cal
  V} = -\int \sigma _{ik}{\dot u}_{ik}\,d{\cal V}=-2\int \mu\, u_{ik}{\dot u}_{ik}d{\cal
  V}
   \end{eqnarray}
which is obtained by scalar multiplication of equation of solid
mechanics, \ref{e2.1}, with $u_i$ and integration over the volume of
seismogenic layer. For our further purpose the field ${\bf u}({\bf
r},t)$ can be conveniently represented in the following separable
form
\begin{eqnarray}
 \label{e2.4}
 {\bf u}({\bf r},t)={\bf a}({\bf r})\,\alpha(t)
 \end{eqnarray}
where ${\bf a}({\bf r})$ is the field of instantaneous
(time-independent) displacements obeying, as follows from
(\ref{e2.2}), to equations
\begin{eqnarray}
  \label{e2.5}
 && \nabla^2 {\bf a}({\bf r})=0,\quad\quad \nabla \cdot {\bf a}({\bf r})=0
 \end{eqnarray}
and $\alpha(t)$ stands for the temporal amplitude of vibrations.
Inserting (\ref{e2.3}) in (\ref{e2.2}) we arrive at equation for
${\alpha}(t)$ having the form of standard equation of linear
oscillations
\begin{eqnarray}
 \label{e2.6}
 && \frac{dE}{dt}=0,\quad E=\frac{M{\dot\alpha}^2}{2}+\frac{K{\alpha}^2}{2}
 \quad\to\quad {\ddot\alpha}+\omega^2\alpha=0,\quad\quad \omega^2=\frac{K}{M},\\
 \label{e2.7}
 && M=\int \rho\, a_i\,a_i\,d{\cal V},\quad\quad
 K=2\int \mu\, a_{ik}\,a_{ik}\,d{\cal V}\quad  \quad a_{ik}=\frac{1}{2}[\nabla_i a_k + \nabla_k
 a_i].
 \end{eqnarray}
 The solenoidal fields of instantaneous material
 displacements in two fundamental modes of nodeless vibrations --
 the spheroidal (normally abbreviated as
 $_0s_\ell$) and the toroidal (abbreviated
 as $_0t_\ell$), are determined by two fundamental (orthogonal and different in
 parity) solutions to the vector Laplace equation of the vector Laplace
 equation built on the general solution to the scalar Laplace equation
 $\nabla^2\chi({\bf r})=0$. In spherical coordinates with fixed polar axis, the solution
 of (\ref{e2.5}) corresponding to nodeless vibrations in the spheroidal mode, ${\bf a}_s$, is given by
 the even parity poloidal (polar) vector field
 and instantaneous displacements in the torsional mode, ${\bf a}_t$, are  by
 the odd parity toroidal (axial) vector field:
 \begin{eqnarray}
 \label{e2.8}
&& {\bf a}_s=\nabla \times \nabla\times\,({\bf
 r}\,\chi),\quad\quad {\bf a}_t=\nabla \times \, ({\bf r}\chi),\\
&&
 \chi({\bf r})=f_\ell({\bf r})P_\ell(\cos\theta),\quad\quad
 f_\ell({\bf r})=[{\cal A}_\ell\,r^\ell+{\cal
  B}_\ell\,r^{-(\ell+1)}].
 \end{eqnarray}
 Henceforth $P_\ell(\cos\theta)$ stands for the Legendre polynomial of multipole
 degree $\ell$ and ${\cal A}_{\ell}$ and ${\cal B}_{\ell}$
 are the arbitrary constants to be eliminated from boundary conditions on the core-crust interface and
 on the star surface. Thus, the problem of computing the frequency
 spectra of both spheroidal and torsional nodeless vibrations is to
 fix the these constants and compute integrals for integral parameters
 of vibrations, that is, the inertia $M$ and the stiffness $K$.

 \subsection{Spheroidal mode}
  The poloidal field of  nodeless instantaneous displacement ${\bf a}_s$ in s-mode
 is irrotational: $\nabla\times {\bf a}_s=0$ (Bastrukov et al 2007b).
 To specify ${\cal A}_{\ell}$ and ${\cal B}_{\ell}$
 we adopt on the core-crust interface, $r=R_c$, the condition
 of impenetrability of seismic perturbation in the core. On the star edge, $r=R$, we impose the condition that
 the radial velocity of material displacements equals the rate of spheroidal distortions of the star
 surface
  \begin{eqnarray}
  \label{e3.1}
  u_r\vert_{r=R_c}=0,\quad\quad {\dot u}_r\vert_{r=R}={\dot R}(t),\quad\quad
  R(t)=R[1+\alpha(t)\,P_\ell(\cos\theta)].
 \end{eqnarray}
 The solution of resultant algebraic equations leads to following
 values of arbitrary constants
\begin{eqnarray}
 \label{e3.2}
  {\cal A}_\ell=\frac{{\cal N}_\ell}{\ell(\ell+1)},\quad
  {\cal B}_\ell=-\frac{{\cal N}_\ell}{\ell(\ell+1)}\,R_c^{2\ell+1},\quad
  {\cal N}_\ell=\frac{R^{\ell+3}}{R^{2\ell+1}-R_c^{2\ell+1}}.
  \label{e2.10}
  \end{eqnarray}
  Tedious but simple calculation of integrals for inertia $M$ and stiffness $K$, given by (\ref{e2.7}),
  with poloidal field ${\bf a}_s$  yields
 \begin{eqnarray}
  \label{e3.3}
  && M_s(\ell,\lambda)
=\frac{4\pi R^5\rho}{\ell(2\ell+1)(1-\lambda^{2\ell+1})}
\left[1+\frac{\ell}{(\ell+1)}\lambda^{2\ell+1}
\right],\\
 \label{e3.4}
&& K_s(\ell,\lambda) =8\pi R^3 \mu
\frac{(\ell-1)(1-\lambda^{2\ell-1})}{\ell(1-\lambda^{2\ell+1})^2}\left[
 1 +
 \frac{\ell(\ell+2)}{\ell^2-1}\,\frac{\lambda^{2\ell-1}(1-\lambda^{2\ell+3})}{(1-\lambda^{2\ell-1})}\right],\\
 \label{e3.5}
 && \lambda=\frac{R_c}{R}=1-h\quad\quad h=\frac{\Delta R}{R}.
 \end{eqnarray}
 The fractional frequency of nodeless spheroidal irrotational shear vibrations
 as a function of multipole degree $\ell$
 is given by
 \begin{eqnarray}
 \label{e3.6}
&&\frac{\omega^2_s(\ell)}{\omega_0^2}=\frac{\nu^2(_0s_\ell)}{\nu_0^2}=\frac{2(2\ell+1)}{(1-\lambda^{2\ell+1})}
\left[\frac{(\ell^2-1)(1-\lambda^{2\ell-1})+\ell(\ell+2)\lambda^{2\ell-1}(1-\lambda^{2\ell+3})}
 {(\ell+1)+\ell\lambda^{2\lambda+1}}\right],\\
  \label{e3.7}
&& \omega_0=\frac{c_t}{R},\quad c_t=\sqrt{\frac{\mu}{\rho}},\quad
\nu_0=\frac{\omega_0}{2\pi},\quad
\nu(_0s_\ell)=\frac{\omega_s(\ell)}{2\pi}.
\end{eqnarray}
 It worth noting that in the limit of zero-size radius of the core,
 $\lambda=(R_c/R)\to 0$, when entire volume of the star sets in
 vibrations, we regain the early obtained spectral formula for global
 nodeless spheroidal nodeless shear vibrations
 $\nu(_0s_\ell)=\nu_0\,[2(2\ell+1)(\ell-1)]^{1/2}$ showing that the
 lowest overtone of the global
 nodeless spheroidal oscillations in the entire volume of the star
 is of quadrupole degree, $\ell=2$ (Bastrukov et al 2002a 2002b; 2007b). In the meantime,
 the lowest overtone of spheroidal vibrations trapped in the crust is of the dipole
 degree, $\ell=1$. This suggests that the dipole overtone can be considered
 as a signature of spheroidal vibrations locked in
 the crust.  In the upper panel of Fig.1 we plot the fractional frequency $\omega_s(\ell)/\omega_0$
 as a function of $h=\Delta R/R$. This picture indicates that dipole vibration can be thought of as, so called,
 Goldstone's soft mode whose most conspicuous feature is that
 the frequency as a function of intrinsic parameter $\lambda$ of oscillating
 system $\omega_s(\ell=1,\lambda)\to 0$, when $\lambda\to 0$.
 In the model under consideration this parameter is given by
 $\lambda=(R_c/R)$. The limit $\lambda=0$ belongs to
 translation displacement of the center-of-mass of the star, not a vibration; this is clearly seen
 from the equation for energy (Hamiltonian) of harmonic oscillations (\ref{e2.6}).
 In the next section we show how the input
 parameters of obtained spectral equation (\ref{e3.6}), namely, the natural unit of
 frequency $\nu_0$ and the depth $h$ of
 seismogenic layer can be extracted from the data on QPOs for SGS.

 \begin{figure}
 \centering{\includegraphics[width=10.cm]{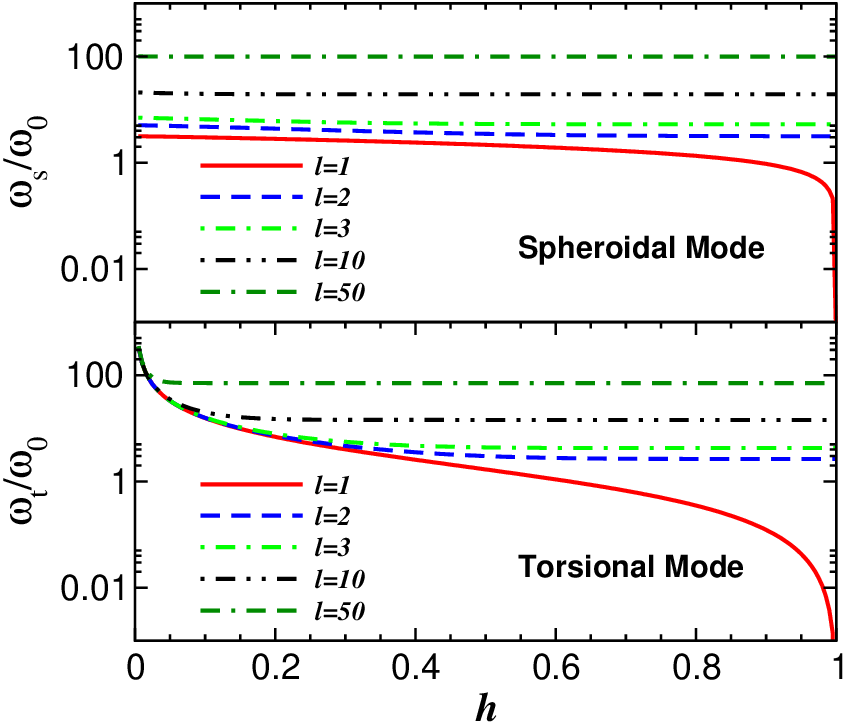}} \caption{
 Fractional frequency of nodeless spheroidal and torsional elastic oscillations as a function of depth
 of seismogenic layer.}
\end{figure}

\subsection{Torsion mode}
  For the torsional oscillations locked in the crust, the constants ${\cal A}_\ell$ and
 ${\cal B}_\ell$ are eliminated from the following boundary conditions
  \begin{eqnarray}
   \label{e4.1}
 && u_\phi\vert_{r=R_c}=0,\quad u_{\phi}\vert_{r=R}=[\mbox{\boldmath $\phi$}_R\times {\bf R}]_\phi, \\
  \label{e4.2}
 && \mbox{\boldmath $\phi$}_R=\alpha(t)\nabla_{\hat{\bf n}} P_\ell(\zeta),\quad\quad
 \nabla_{\hat{\bf n}}=\left(0,\frac{\partial }{\partial
 \theta},\frac{1}{\sin\theta}\frac{\partial }
 {\partial \phi}\right).
 \end{eqnarray}
 First is the no-slip condition on the core-crust interface, $r=R_c$,
 implying that the amplitude of differentially rotational oscillations
 is gradually decreasing from the surface to the core.
 The boundary condition on the star surface, $r=R$,
 is dictated by symmetry of the general toroidal solution of the vector Laplace equation.
 The support of this last boundary condition lends further considerations showing that
 it leads to correct expression for the moment of inertia of a rigidly rotating star.
 The resultant algebraic equations steaming from above boundary conditions
 lead to
  \begin{eqnarray}
  \label{e4.3}
 {\cal A}_\ell={\cal N}_\ell,\quad {\cal B}_{\ell}=-{\cal
 N}_\ell\,R_c^{2\ell+1},\quad\quad {\cal
 N}_\ell=\frac{R^{\ell+2}}{R^{2\ell+1}-R_c^{2\ell+1}}.
 \end{eqnarray}
 Tedious calculation of integrals for $M_t$ and $K_t$ leads to
 \begin{eqnarray}
  \label{e4.4}
 && M_t=\frac{4\pi\ell(\ell+1)}{(2\ell+1)(2\ell+3)}\frac{\rho R^5}{(1-\lambda^{2\ell+1})^2}\times\\
  \nonumber
 &&\left[1- (2\ell+3)\lambda^{2\ell+1}+
\frac{ (2\ell+1)^2}{2\ell-1}
\lambda^{2\ell+3}-\frac{2\ell+3}{2\ell-1}\lambda^{2(2\ell+1)}\right],\\[0.5cm]
 \label{e4.5}
 && K_t=\frac{4\pi\ell(\ell^2-1)}{2\ell+1}\,\frac{\mu R^{3}}{(1-\lambda^{2\ell+1})}
 \left[1+\frac{(\ell+2)}{(\ell-1)}\lambda^{2\ell+1}\right]\\
 && \lambda=\frac{R_c}{R}=1-h\quad h=\frac{\Delta R}{R}.
\end{eqnarray}
In the limit of zero-size radius of the core, $\lambda=(R_c/R)\to
 0$, corresponding to torsional oscillations in the entire volume of
the star  we regain the early obtained spectral formula for the
global nodeless torsional elastic vibrations
$\nu(_0t_\ell)=\nu_0\,[(2\ell+3)(2\ell-1)]^{1/2}$ showing that in
case of global torsional oscillations the lowest overtone is of
quadrupole degree (Bastrukov et al 2002a; 2002b; 2007a; 2007b).
However, this is not the case when we consider torsional nodeless
oscillations locked in the seismogening layer of finite depth
$\Delta R=R-R_c$. For $\ell=1$, equations (\ref{e4.4}) and
(\ref{e4.5}) are reduced to
\begin{eqnarray}
  \label{e4.6}
 && M_t(\ell=1,\lambda)=\frac{8\pi\,\rho\,R^5}{15(1-\lambda^3)^2}\left[1-5\lambda^3+9\lambda^5-5\lambda^6\right],\\
 \label{e4.7}
 && K_t(\ell,\lambda)=8\pi\,\mu \,R^3\,\frac{\lambda^3}{(1-\lambda^3)},\\
  \label{e4.8}
 && \omega_t^2(\ell=1,\lambda)=\omega_0^2\,
 \frac{15\lambda^3(1-\lambda^3)}{(1-\lambda)^3(1+3\lambda+6\lambda^2+5\lambda^3)}\quad  0\leq \lambda <  1.
\end{eqnarray}
 In the limit zeroth core, $\lambda\to 0$, the stiffness $K_t\to 0$ and the mass
 parameter getting the form of moment of inertia of absolutely rigid
 solid star of mass ${\cal M}$ and radius $R$:
 $M_t(\ell=1,\lambda=0)=(2/5){\cal M}R^2$. This consideration again shows that the dipole vibration
 exhibits features of the Goldstone soft mode owing its emergence to the trapping of torsional shear
 oscillations in the peripheral crust of finite depth.

\begin{figure}
\centering{\includegraphics[width=10.cm]{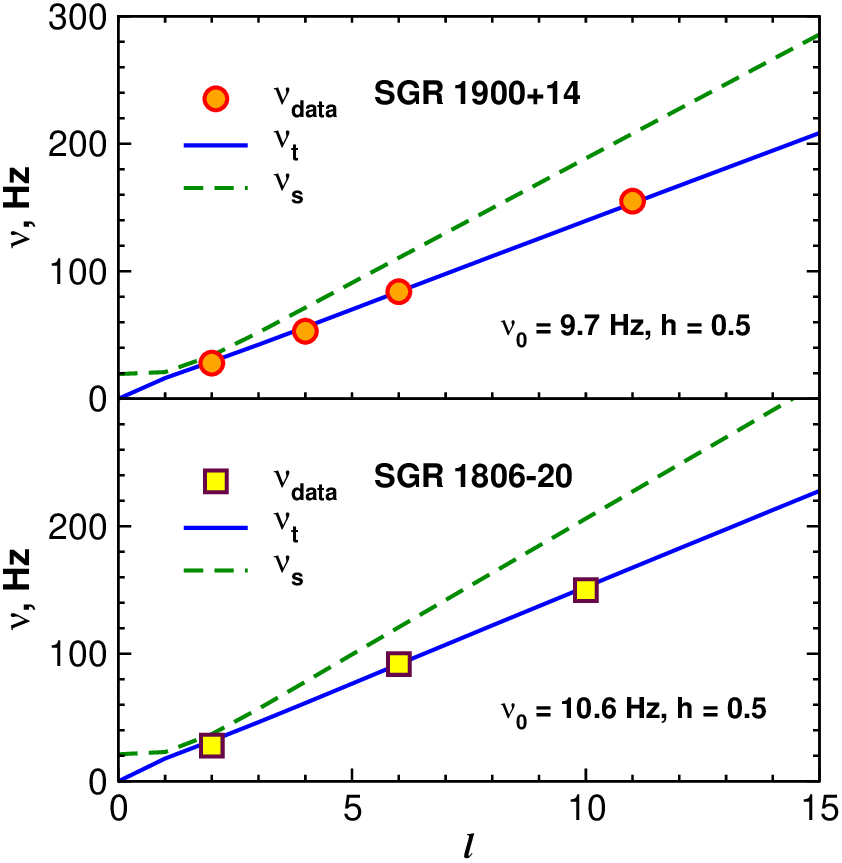}} \caption{
 Theoretical curves for the frequency of spheroidal (dashed) and torsional (solid) nodeless elastic
 oscillations computed with aid of spectral formulae for frequency of spheroidal, eq.(15), and
 torsional,
 eqs. (26)-(27),
 modes as functions of multipole degree in juxtaposition with data (symbols) on QPOs for SGR 1900+14
 and for SGR 1806-20. The modal identification is taken from (Samuelsson and Andersson 2007b;  Watts \& Strohmayer 2007).}
\end{figure}

 The general spectral equation for the
 fractional frequency of nodeless torsional oscillations of arbitrary multipole degree
 $\ell$, computed with aid of equations (\ref{e4.4}) and
 (\ref{e4.5}), can be presented in the following analytic form
\begin{eqnarray}
 \label{e4.9}
 &&\frac{\omega_t^2(\ell)}{\omega_0^2}=\frac{\nu^2(_0t_\ell)}{\nu_0^2}=[(\ell+2)(\ell-1)]\,p_\ell(\nu_0,\lambda)\\
 \label{e4.10}
 && p_\ell(\nu_0,\lambda)=4\left[1-\frac{1}{2(\ell+2)}\right]\left[1+\frac{1}{2(\ell-1)}\right]
 (1-\lambda^{2\ell+1})\times\\
\nonumber
 &&\left\{1-
\frac{\ell-\lambda^{2\ell+1}[(\ell+2)+(2\ell-1)(2\ell+3)-(2\ell+1)^2\lambda^2+(2\ell+3)\lambda^{2\ell+1}]
}{\quad\,
(2\ell-1)-\lambda^{2\ell+1}[(2\ell-1)(2\ell+3)-(2\ell+1)^2\lambda^2+(2\ell+3){\lambda}^{2\ell+1}]
} \right\}.
\end{eqnarray}
The practical usefulness of such representation is extensively
discussed in (Bastrukov et al 2007a). The fractional frequency as a
function of $h=\Delta R/R$ is pictured in down panel of Fig.1 which
show that the lowest overtone of torsional vibrations trapped in the
crust is of dipole degree and that dipole overtone of differentially
rotational vibrations of the crust against core possesses properties
of the Goldstone's soft mode.

\section{Application to SGR 1900+14 and SGR 1806-20}

 The obtained spectral formulae (15) and (26)-(27) describe the frequencies of both spheroidal and torsional
 nodeless oscillations
 as functions of the multipole degree $\ell$. The natural unit of frequency $\nu_0$ of shear elastic vibrations
 and the fractional depth $h$ of peripheral
 seismogenic layer are input parameters carrying information about material properties of
 neutron star matter (density, shear modulus) and geometrical sizes of star and seismoactive zone.
 Considering the observation data for SGR 1900+14 and SGR 1806-20, we demonstrate here how
 the obtained spectral equations can be used to eliminate some uncertainties
 in identification of QPOs.

 \begin{figure}
 \centering{\includegraphics[width=9.cm]{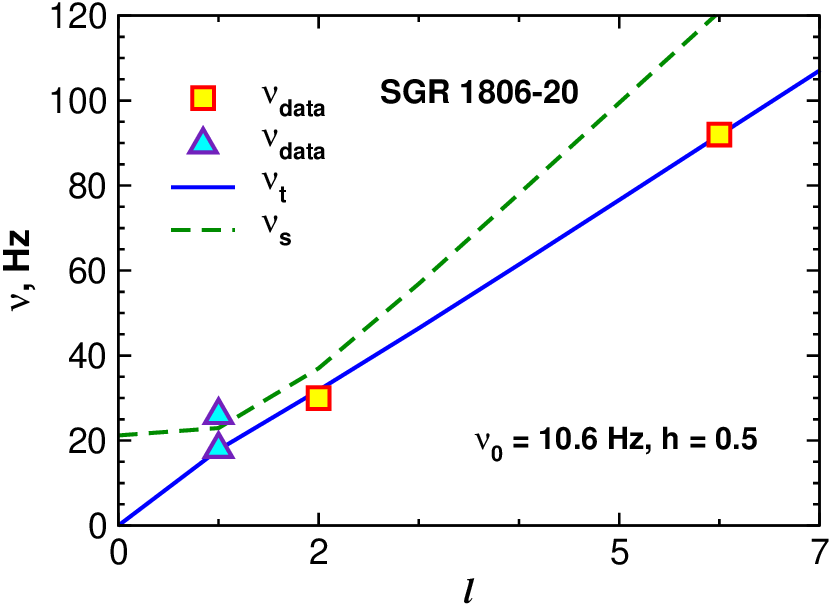}}
 \caption{
 Theoretical predictions for the frequency of spheroidal
 (dashed) and torsional  (solid) nodeless elastic
 oscillations as a function of multipole degree $\ell$ in juxtaposition with data (symbols) on QPOs
 SGR 1806-20. The identification of QPOs pictured by squares is taken from (Samuelsson \& Andersson 2007b;
 Watts \& Strohmayer 2007).
 Based on the results of this latter work, our calculations suggest that low frequency QPOs discovered in
 (Israel et al 2005), that are pictured by triangles, can be identified as dipole toroidal and dipole spheroidal
 nodeless vibration, respectively: $\nu(_0t_1)=18$ Hz and
 $\nu(_0s_1)=26$ Hz.
 }
 \end{figure}

 First, we examine the agreement of obtained  spectral formula (\ref{e4.9})-(\ref{e4.10}) for torsion mode
 with identification of the QPOs frequencies from interval $30\leq \nu\leq
 200$ Hz with frequencies of nodeless torsional vibrations of
 multipole degree $\ell$ from interval $2\leq \ell \leq 12$ suggested
 in works (Samuelsson \& Andersson 2007a; 2007b).
 In so doing we use the proposed in these latter works
 identification of QPOs in SGR 1900+14 data [namely, $\nu(_0t_2)=28$ Hz; $\nu(_0t_4)=53$; Hz
 $\nu(_0t_6)=84$ Hz, $\nu(_0t_{11})=155$ Hz borrowed from Table 1 of paper
 (Samuelsson \& Andersson 2007b)] as reference points and vary parameters $\nu_0$ and $h$ entering
 our spectral formula (15)
 for torsional mode so as to attain the best fit of these points. The result of this
 procedure is shown in upper panel of Fig.2 by solid line.
 Then, making use of the fixed in the above manner parameters $\nu_0$ and $h$, we compute
 (with the aid of spectral formula (\ref{e3.6})) the frequency of spheroidal
 mode $\nu(_0s_\ell)$. The application of this procedure to  modal analysis of QPOs data for
 SGR 1806-20 is pictured in down panel of Fig.2. Based on proposed in the above mentioned paper identification
 of the following points $\nu(_0t_2)=30$ Hz;  $\nu(_0t_6)=92$ Hz and  $\nu(_0t_{10})=150$ Hz we
 extract parameters $\nu_0$ and $h$ entering in our spectral formulae for torsional
 mode, equations (\ref{e4.9})-(\ref{e4.10}).

 In Fig.3 we highlight by triangles two non-identified before
 low-frequency QPOs in data for SGR 1806-20 (Israel et al
 2005), namely $\nu=18$ Hz and $\nu=26$ Hz.
 The extrapolation to low-frequency region of spectral formulae (13)
 and (15) leads us to conclude that the latter low-frequency points can be interpreted as manifestation of
 the dipole toroidal, $_0t_1$, and the dipole
 spheroidal, $_0s_1$, overtones of nodeless shear elastic vibrations, respectively.

 As to high-frequency points $\nu=626$ Hz and $\nu=1840$ Hz in
 data for SGR 1806-20 is concerned, the lack of
 observational data makes the above scheme of identification less effective.
 Putting these points on spectral curve for
 torsional mode $\nu(_0t_\ell)$ extended to very high value of
 $\ell$, we get $\nu(_0t_{\ell=42})=625$ Hz and
 $\nu(_0t_{\ell=122})=1840$. However, if one puts these points on the
 spectral curve for spheroidal mode $\nu(_0s_\ell)$,
 we obtain the following
 identification $\nu(_0s_{\ell=30})=625$ Hz and
 $\nu(_0t_{\ell=87})=1840$. The last case may be more
 favorable because the
 higher multipole degree of oscillations the less their lifetime (Bastrukov et al 2007b), and
 low-$\ell$ overtones have, therefore, more chances for surviving.
 But one must admit that this viewpoint is highly questionable
 and should be thought of as
 suggestive, not conclusive.

\section{Summary}

 The exact spectral formulae, which has been obtained here within the
 framework of
 Newtonian, non-relativistic, solid-mechanical theory of seismic vibration for the first time,
 are interesting in its own right from the viewpoint of general theoretical
 seismology (e.g. Lay, Wallace 1995) because
 they can be utilized in the study of seismic vibrations of more wider class of solid celestial objects such
 as Earth-like planets.
 One of the remarkable findings of our investigation is that the dipole overtones
 of nodeless elastic shear vibrations trapped in the finite-depth crust of seismically active neutron star
 possess properties of Goldstone soft modes.
 It is shown that obtained spectral equations
 are consistent with the existence treatment of low-frequency QPOs in the X-ray luminosity of flares
 SGR 1900+14 and  SGR 1806-20 as caused by quake-induced torsional nodeless vibrations
 (Samuelsson \& Andersson 2007a; 2007b).
 What is newly disclosed here is that previously non-identified low-frequency QPOs in data for SGR 1806-20
 can be attributed to nodeless dipole torsional and spheroidal
 vibrations, namely, $\nu(_0t_1)=18\,\mbox{Hz}$ and $\nu(_0s_1)=26\,\mbox{Hz}$.

\section{Acknowledgements}

 This work is partly supported by NSC of Taiwan,
 under grants NSC-
 96-2811-M-007-012 and NSC-96-2628-M-007-012-MY3.
 The authors are grateful to Dr. Judith Bunder (UNSW, Sydney) for critical reading of the manuscript.
 Also, we are indebted to anonymous referee for several valuable remarks and questions clarifying understanding
 of problems touched upon in this work.

\end{document}